\documentclass{prop2015}
\usepackage[english]{babel}
\usepackage{amssymb,slashed,amsmath}

\newcommand{\zbar}{\bar{z}^{\bar{1}}}
\newcommand{\gone}[1]{g_{1\bar{1}#1}}
\newcommand{\kinv}[2]{(K_{0,\bar{#1}{#2}})^{-1}}

\def\rme{{\rm e}}

\def\K{K\"{a}hler}

\def\aD3{{\overline {\rm D3}}}
\def\be{\begin{equation}}
\def\ee{\end{equation}}
\def\ba{\begin{array}}
\def\ea{\end{array}}
\def\bea{\begin{eqnarray}}
\def\eea{\end{eqnarray}}
\def\ib{{\bar \imath}}
\def\jb{{\bar \jmath}}
\def\zb{{\bar z}}
\def\b1{{\bar1}}
\def\K{K{\"a}hler}
\def\w{\wedge}
\newcommand{\Ls}[1]{\mathcal{L}_{\rm #1}}
\newcommand{\lp}{\left(}
\newcommand{\ls}{\left[}
\newcommand{\rp}{\right)}
\newcommand{\rs}{\right]}
\newcommand{\nn}{\nonumber}

\title{The general de Sitter supergravity component action}
\author[M. Schillo]{Marjorie Schillo\inst{1}}
\author[E. van der Woerd]{Ellen van der Woerd\inst{1}}
\author[T. Wrase]{Timm Wrase\inst{2}
\footnote{E-mail:~\textsf{marjorie@itf.fys.kuleuven.be}, \textsf{ellen@itf.fys.kuleuven.be}, \textsf{timm.wrase@tuwien.ac.at}}
}
\address[1]{Institute for Theoretical Physics, University of Leuven, 3001 Leuven, Belgium}
\address[2]{Institute for Theoretical Physics, TU Wien, A-1040 Vienna, Austria}
\begin{abstract}
 In this paper we review the appearance and utility of a nilpotent chiral multiplet in the context of supergravity, string theory and cosmology. \\
Coupling a nilpotent chiral superfield to supergravity, one obtains what is called \emph{pure dS supergravity}, a supergravity theory without scalar degrees of freedom that naturally has de Sitter (dS) solutions, and in which supersymmetry is non-linearly realized. We extend previous results that couple this dS supergravity to chiral and vector multiplets and derive the most general supergravity action for a single nilpotent chiral multiplet coupled to supergravity and an arbitrary number of chiral and vector multiplets.

\emph{Based in part on the plenary talk given by T. W. at ``The String Theory Universe'', 21st European String Workshop, Leuven, September 7-11, 2015.}
\end{abstract}
\shortabstract
\begin{document}
\maketitle

\section{Introduction}
More than forty years ago Volkov and Akulov (VA) \cite{Volkov:1972jx, Volkov:1973ix} speculated that a neutrino in the standard model of particle physics could be the massless goldstino, $\chi$, that arises when supersymmetry is spontaneously broken. They derived an explicit action for the goldstino that is invariant under non-linearly realized supersymmetry transformation. This VA action is given by
\be\label{eq:VAaction}
S_{\rm VA} = M^4 \int E^0 \w E^1 \w E^2 \w E^3\,,\quad E^\mu = dx^\mu + \bar \chi \gamma^\mu d\chi\,.
\ee
The non-linear supersymmetry transformation under which this action is invariant is
\be\label{eq:VAtrafo}
\delta_\zeta \chi = \zeta + (\bar \chi \gamma^\mu \zeta) \ \partial_\mu \chi\,,
\ee
where $\zeta$ is a constant spinor. 

With the discovery of neutrino oscillations it became clear at the end of the last century that the standard model neutrinos are not massless. Nevertheless, the VA action is interesting on theoretical grounds and might still play some other role in the description of our universe. The work \cite{Rocek:1978nb, Ivanov:1978mx, Lindstrom:1979kq, Casalbuoni:1988xh, Komargodski:2009rz} following VA established that the goldstino, $\chi$, can be packaged into a nilpotent chiral multiplet in 4d $\mathcal{N}=1$ supersymmetry.\footnote{It was also shown in \cite{Kuzenko:2010ef, Kuzenko:2011tj} that there is  a unique action for the goldstino up to a non-linear field redefinition.} If we denote a chiral superfield by $S$ and demand that it squares to zero, $S^2=0$, then we find the following equations:
\bea
S^2 &=& (s+\sqrt{2}\theta \chi + \theta^2 F)^2 \cr
&=& s^2 + 2\sqrt{2} s \ \theta \chi + \theta^2 (2s F -\bar \chi P_L \chi)\,.
\eea
The three equations $s^2= s P_L \chi = 2s F -\bar \chi P_L \chi=0$ are all simultaneously solved by 
\be\label{eq:s}
s=\frac{\bar \chi P_L \chi}{2 F}. 
\ee
So we see that the nilpotent chiral multiplet, $S$, has only fermionic degrees of freedom since the scalar is given by a fermion bilinear.

As we will review in section \ref{sec:review}, this nilpotent chiral multiplet has recently been introduced in the cosmological context in \cite{Antoniadis:2014oya} (see also \cite{AlvarezGaume:2010rt, AlvarezGaume:2011xv, AlvarezGaume:2011db, Achucarro:2012hg} for earlier work). This has triggered a plethora of successive papers 
\begin{itemize}
\item that use the nilpotent multiplet and its interesting features in the context of cosmological model building \cite{Ferrara:2014ima, Ferrara:2014fqa, Buchmuller:2014pla,Ferrara:2014cca, Ferrara:2014kva, Kallosh:2014via, Dall'Agata:2014oka, Kallosh:2014hxa, Kallosh:2015lwa, Linde:2015uga, Carrasco:2015uma, Kahn:2015mla, Carrasco:2015rva, Scalisi:2015qga, Carrasco:2015pla, Dudas:2015eha}, 
\item study its connection to string theory \cite{McGuirk:2012sb, Ferrara:2014kva, Kallosh:2014wsa, Bergshoeff:2015jxa, Kallosh:2015nia},
\item and study the supergravity action in the presence of a nilpotent field or even more general constrained multiplets \cite{Dudas:2015eha, Bergshoeff:2015tra, Hasegawa:2015bza, Kuzenko:2015yxa, Antoniadis:2015ala, Kallosh:2015sea, Kallosh:2015tea, Dall'Agata:2015zla}.
\end{itemize} 
In particular, if one couples only a single nilpotent chiral multiplet to supergravity, one obtains the so called \emph{pure dS supergravity}, a supergravity theory without scalar fields that naturally gives rise to dS solutions! This was first shown in \cite{Antoniadis:2014oya} where the bosonic action was studied. Then in \cite{Ferrara:2014kva} the superconformal action using superfields was spelled out. However, the component action, including fermions, for this \emph{pure dS supergravity} was worked out only recently in \cite{Bergshoeff:2015tra, Hasegawa:2015bza}. The component action for \emph{dS supergravity} coupled to one chiral multiplet was worked out in \cite{Hasegawa:2015bza}. This was extended in \cite{Kallosh:2015sea, Kallosh:2015tea}, where the coupling to an arbitrary number of chiral and vector multiplets was worked out under the single assumption that the {\K} potential only depends on $s \bar s$ and has no terms linear in $s$ or $\bar s$. After reviewing the current literature involving the nilpotent chiral multiplet in section \ref{sec:review}, we extend the result of \cite{Kallosh:2015tea} in section \ref{sec:action} and work out the most general component action for \emph{dS supergravity} coupled to an arbitrary number of chiral and vector multiplets. We conclude in section \ref{sec:conclusion}. In two appendices we collect useful but lengthy results related to the derivation of the action in section \ref{sec:action}. 

\section{Review of the nilpotent multiplet}\label{sec:review}

\subsection{dS vacua and inflation with a nilpotent chiral superfield}\label{sec:dS}
In 2014, the nilpotent chiral multiplet was used in the cosmological context in \cite{Antoniadis:2014oya}. In that paper the authors showed (among other things) that it is trivial to construct dS vacua in supergravity models with a nilpotent chiral multiplet. In particular, consider the following {\K} potential and superpotential:
\be
K=-3 \log(1-s \bar s) =3s \bar s\,,\qquad W = f s+W_0\,,
\ee
where $s$ is nilpotent so that $s^2=0$. Now one can calculate the scalar potential via the usual formula and set $s=0$ \footnote{We are not aware of a mechanism that could give a vacuum expectation value to the fermion bilinear $s$.} in order to get the bosonic action:
\bea \label{eq:ftermbosonic}
V&=&e^K\lp K^{s\bar s} D_s W \overline{D_s W} -3|W|^2\rp = \frac13 |f|^2 -3|W_0|^2\,,\cr
m_{3/2}^2 &=&|W_0|^2\,,
\eea
with $D_S W= \partial_S W + W \partial_S K$. Thus, in this model the cosmological constant is decoupled from the gravitino mass, $m_{3/2}$, and it is trivial to find dS vacua, i.e. to arrange for $V>0$. Note that these dS vacua arise without any scalar degrees of freedom! The reason that this is possible, despite the no-go theorems in \cite{Pilch:1984aw, Lukierski:1984it}, is that supersymmetry is spontaneously broken due to the nilpotent chiral multiplet.

Additionally, \cite{Antoniadis:2014oya} discussed the role of the nilpotent chiral multiplet in the context of the Starobinsky model of inflation \cite{Starobinsky:1980te}. In this model there exists a classical duality between a scalar field coupled to gravity and an $R+R^2$ action, where $R^2$ denotes a higher derivative correction to the Einstein-Hilbert Lagrangian density $R$. This model was embedded into (higher-derivative) supergravity in \cite{Cecotti:1987sa, Cecotti:1987qe, Ferrara:2013wka}. In the so-called `old minimal version of supergravity', this supergravity generalization of the $R+R^2$ model is dual to the usual Einstein-Hilbert action coupled to two chiral multiplets $T$ and $S$. The field $T$ contains the inflaton and \cite{Antoniadis:2014oya} studies the case in which $S$ is a nilpotent chiral multiplet so that $S^2=0$. They find that in the dual $R+R^2$ model the nilpotency of $S$ translates into a nilpotency condition for the chiral curvature field, $\mathcal{R}$, that is now constrained to $\mathcal{R}^2=0$. 

These models were extended and new ones were derived and studied in \cite{Ferrara:2014ima, Ferrara:2014fqa, Ferrara:2014cca, Ferrara:2015ela, Dudas:2015eha, Ferrara:2015gta, Antoniadis:2015ala, Hasegawa:2015era}. Among other things, models were discovered in which the chiral curvature field, $\mathcal{R}$, is not really nilpotent but rather satisfies the constraint $(\mathcal{R}-\lambda)^2=0$. For an in-depth discussion of the Starobinsky model of inflation \cite{Starobinsky:1980te} and higher-derivative supergravity in the context of a nilpotent chiral curvature superfield see the nice review \cite{Ferrara:2015cwa}. \footnote{See also \cite{Ferrara:2015exa} which additionally reviews the partial breaking from $\mathcal{N}=2$ to $\mathcal{N}=1$ due to nilpotent superfields.}

It was pointed out in \cite{Ferrara:2014kva} that the nilpotent chiral superfield should simplify all inflationary models that require a so-called `stabilizer' chiral multiplet in addition to the chiral multiplet that contains the inflaton. One can simply replace this stabilizer field by a nilpotent field which effectively removes its scalar component and also usually simplifies the {\K} potential. Furthermore, it is possible to use the nilpotent field to obtain dS vacua at the end of inflation as was shown in newly constructed models in \cite{Kallosh:2014via}.
 
The nilpotent chiral multiplet has been extensively used in the context of $\alpha$-attractors. These are inflationary models in supergravity that were first proposed in 2013 in \cite{Ferrara:2013rsa} (see also \cite{Kallosh:2013hoa, Kallosh:2013yoa}). These inflationary models can accommodate a wide variety of possible cosmological measurements that will be made in the near future. In the papers \cite{Kallosh:2014hxa, Kallosh:2015lwa, Linde:2015uga, Carrasco:2015uma, Carrasco:2015rva, Scalisi:2015qga, Carrasco:2015pla} a variety of simple models have been constructed. These can not only accommodate any possible measurement for the tensor-to-scalar ratio, $r$, and the spectral index, $n_s$, but also allow for a tunably small cosmological constant at the end of inflation, as well as an arbitrary SUSY breaking scale. The paper \cite{Carrasco:2015rva} also studies the initial conditions in these models and finds that inflation arises naturally.

Additional work with the nilpotent chiral multiplet has resulted in the construction of other inflationary models and dS vacua, as well as further studies of the resulting supergravity action. In \cite{Buchmuller:2014pla} supersymmetry breaking in large field models of inflation was studied and one of the models featured the nilpotent chiral multiplet. A novel class of inflationary models in supergravity without a sgoldstino was constructed in \cite{Dall'Agata:2014oka} by using a single nilpotent chiral multiplet. The paper \cite{Kahn:2015mla} constructed the minimal effective field theory of supersymmetric inflation using not only a nilpotent chiral superfield but also a constrained real superfield. In the paper \cite{Kuzenko:2015yxa} it was shown that a complex linear goldstino superfield coupled to supergravity can likewise lead to dS vacua and the relation of this setup to a nilpotent chiral multiplet was discussed in detail. The paper \cite{Dall'Agata:2015zla} constructed the minimal supergravity model where the nilpotent chiral goldstino superfield is coupled to a chiral matter superfield, in particular, the two chiral multiplets $S$ and $Y$ were constrained to satisfy $S^2 = S Y =0$.

In summary, the nilpotent chiral multiplet renders the construction of dS vacua in supergravity trivial, making it an invaluable ingredient in cosmological model building. Furthermore, the construction of inflationary models can often simplify and the models can  easily accommodate different features of our universe if one includes a nilpotent chiral multiplet. Since the nilpotent chiral multiplet and its extensions have only started to play a role in cosmological model building very recently, we can expect many interesting developments in the future. 

\subsection{The nilpotent chiral superfield in string theory}\label{sec:string}
The nilpotent chiral superfield also appears in string compactifications. As we will review here, this has been explicitly worked out for the case of Kachru, Kallosh, Linde and Trivedi (KKLT) type compactifications of type IIB string theory \cite{Kachru:2003aw}. In this setting the complex structure moduli and the dilaton are stabilized by $H_3$ and $F_3$ fluxes following Giddings, Kachru and Polchinski (GKP) \cite{Giddings:2001yu}. The single {\K} modulus $T$ that controls the overall volume of the internal warped Calabi-Yau manifold remains a flat direction. After including non-perturbative effects from Euclidean D3-branes or gaugino condensation on a stack of D7-branes, one has the following 4d $\mathcal{N}=1$ effective description in terms of a {\K} potential and superpotential:
\be\label{eq:KKLTKW}
K=- 3 \ln(T+\bar{T})\,,\qquad W = W_0 + A \rme^{-aT}\,.
\ee
Here, $W_0$, $A$ and $a$ are constants and the resulting scalar potential,
\be\label{eq:VKKLT}
V_{\rm KKLT}=e^K\lp K^{T\bar T} D_T W \overline{D_T W} -3|W|^2\rp\,,
\ee
has a supersymmetric AdS vacuum in which all moduli have a positive mass squared. In order to obtain a metastable dS minimum, KKLT added to the scalar potential a positive term that arises from one or more anti-D3-branes:
\be\label{eq:Vuplift}
V= V_{KKLT} + V_{\aD3} = V_{KKLT} + \frac{\mu^4}{(T+\bar T)^3}\,.
\ee
Naively, it seems that this term breaks supersymmetry explicitly, however, it follows, for example, from the work of Kachru, Pearson and Verlinde \cite{Kachru:2002gs}, that this dS vacuum can decay to a supersymmetric ground state and therefore supersymmetry should be broken spontaneously. Nevertheless, until last year it was not known how to modify the {\K} potential and superpotential in equation \eqref{eq:KKLTKW} to get the extra contribution from the anti-D3-brane in equation \eqref{eq:Vuplift}.

It was shown by Ferrara, Kallosh and Linde \cite{Ferrara:2014kva} that using a nilpotent superfield, $S$, one can get the anti-D3-brane uplift term. In particular, by making the following $s$ dependent modification to the {\K} potential and superpotential in equation \eqref{eq:KKLTKW},
\be
K=- 3 \ln(T+\bar{T})+s\bar s\,,\qquad W = W_0 + A \rme^{-aT}+\mu^2 s\,,
\ee
one finds exactly the scalar potential in equation \eqref{eq:Vuplift}:
\bea
V&=& e^K\lp K^{T\bar T} D_T W \overline{D_T W}+K^{s\bar s} D_s W \overline{D_s W} -3|W|^2\rp\Big|_{s=\bar s=0} \cr
&=& V_{KKLT} + \frac{\mu^4}{(T+\bar T)^3}\,.
\eea
It was shown in KKLMMT \cite{Kachru:2003sx} that, in the presence of warping, the contribution of the anti-D3-brane is modified.  In this case the uplift term only has a quadratic dependence on the volume modulus, so that the scalar potential is given by:
\be\label{eq:Vupliftw}
V= V_{KKLT} + V_{\aD3, {\rm warped}} = V_{KKLT} + \frac{\mu^4}{3(T+\bar T)^2}\,.
\ee
This scalar potential can also be reproduced from a {\K} potential and superpotential, if one uses a nilpotent superfield \cite{Ferrara:2014kva}. In particular, the following $K$ and $W$ give rise to the scalar potential in equation \eqref{eq:Vupliftw}:
\be
K=- 3 \ln(T+\bar{T} - s\bar s) \,,\qquad W = W_0 + A \rme^{-aT}+\mu^2 s\,.
\ee
This ability to reproduce the desired scalar potentials  suggests a connection between the nilpotent supermultiplet and the anti-D3-brane uplift. 

This connection was made more precise in \cite{Ferrara:2014kva} where the authors recalled that the fermionic action for a Dp-brane is closely related to the VA action that describes the nilpotent chiral superfield. In particular, in the standard $\kappa$-symmetry fixing of \cite{Aganagic:1996nn, Aganagic:1996pe} the fermionic part of the Chern-Simons term vanishes and one finds that the fermionic part of the DBI action takes the form of the VA action \cite{Kallosh:1997aw, Bergshoeff:2013pia}. Thus, it seems very likely that the spontaneous breaking of supersymmetry by Dp-branes is reflected in the 4d supergravity theory by the presence of a nilpotent chiral multiplet.

From the results above it appears that there is no difference between a brane and an anti-brane, since they only differ in the sign of the Chern-Simons term, which vanished. However, in the KKLT setup a brane preserves supersymmetry and does not uplift the AdS vacuum while an anti-D3-brane breaks supersymmetry and leads to the uplift term. This issue was addressed in \cite{Kallosh:2014wsa}, where it was recalled that the KKLT construction involves an $O3$ orientifold projection that is not compatible with the standard $\kappa$-symmetry gauge fixing that was used in  \cite{Aganagic:1996nn, Aganagic:1996pe}. The orientifold truncation leads for a D3-brane that sits on top of $O3^-$ plane to a vanishing action and for an anti-D3-brane on top of an $O3^-$ plane to an action without $\kappa$-symmetry.\footnote{This follows from the T-dual analysis for 9-branes in \cite{Bergshoeff:1999bx}.} From string theory considerations we know the spectrum \cite{Sugimoto:1999tx, Antoniadis:1999xk, Uranga:1999ib}, and for the case of a single D3-brane the orientifold projection removes all world volume degrees of freedom, i.e. the scalars, vector and fermions, which is consistent with the vanishing of the action in this case. For an anti-D3-brane one is left with only fermions and one can study the purely fermionic action in full detail. One finds that this world volume action for an anti-D3-brane on top of an O3$^-$-plane is a generalization of the standard 4d $\mathcal{N}=1$ VA action (cf. equation \eqref{eq:VAaction}). In $\mathcal{N}=1$ language one has four nilpotent chiral supermultiplets, containing a total of four 4d spinors, that spontaneously break the 4d $\mathcal{N}=4$ supersymmetry preserved by the orientifold projection \cite{Kallosh:2014wsa}. 

This analysis was extend in \cite{Bergshoeff:2015jxa}, where an anti-D3-brane on top of an orientifold plane in a GKP background \cite{Giddings:2001yu} was analyzed (see also \cite{McGuirk:2012sb} for earlier work in this direction). In such a background the action for a single anti-D3-brane is only known to quadratic order in fermions, which is sufficient to show that three of the four fermions get a mass and only one fermion remains massless. This massless fermion spontaneously breaks the linearly realized $\mathcal{N}=1$ supersymmetry that was preserved by the GKP background. Since the action for the anti-D3-brane in this background is only known to quadratic order, it was not possible in \cite{Bergshoeff:2015jxa} to show that the action for this fermion is the VA action to all orders, however, this follows from the uniqueness of the VA action (see \cite{Kuzenko:2011tj} and references therein).

An important step in the KKLT construction is that the anti-D3-brane sits in a highly warped region, i.e. at the bottom of a throat. So one might ask whether it is possible to place the above system of an anti-D3-brane on top of an O3$^-$-plane at the bottom of a throat. This question was addressed by Kallosh, Quevedo and Uranga in \cite{Kallosh:2015nia}. They find that for the canonical example of the Klebanov-Strassler throat \cite{Klebanov:2000hb} it is not possible to have an orientifold projection that leads to O3-planes. However, they explicitly constructed another throat that allows for an orientifold plane that is localized at the bottom of the throat. Placing an anti-D3-brane on top of this O3$^-$ then leads to the redshifted anti-D3-brane uplifing term that one expects in the KKLT construction.  This construction of the uplift term, as well as the direct connection between the anti-D3-brane world volume fermion and the VA action, makes the connection to the nilpotent chiral supermultiplet fully explicit. 

Additionally, the authors of \cite{Kallosh:2015nia} noted that one can introduce O7-planes in the Klebanov-Strassler throat. The world-volume action of an anti-D3-brane placed at the bottom of this throat contains two complex scalars and two fermions. Both scalars and one fermion will receive a mass if one turns on imaginary self-dual background fluxes, and the remaining massless fermion will again have the VA action and spontaneously break supersymmetry. This provides another explicit connection between the anti-D3-brane uplift in the KKLT string theory scenario and the nilpotent chiral multiplet. Since (anti-) branes break supersymmetry spontaneously it is expected that many more such connections between string compactifications and the nilpotent chiral multiplet will be discovered in the future.

\subsection{The component action for dS supergravity}
We have seen above in subsection \ref{sec:dS} that the nilpotent chiral multiplet has led to substantial progress in cosmological model building in supergravity. For the construction of inflationary models and dS vacua, we in principle only need the bosonic action in the presence of a nilpotent chiral superfield. As we reviewed above equation \eqref{eq:ftermbosonic},  this bosonic action is given by dropping all terms quadratic in $s$ and $\bar{s}$ and setting $s=\bar{s}=0$ after calculating the F-term scalar potential. However, at the end of inflation we need to reheat the universe which requires a particle physics sector and a thorough understanding of the fermionic terms in the supergravity action. Thus, it is important to derive the full supergravity action for the nilpotent multiplet coupled to regular chiral and vector multiplets. Such an action might also be interesting purely from a particle physics point of view. In this subsection we review the status of deriving such an action and in the next section we will fill in a gap in the existing literature and derive the most general component action for one nilpotent chiral multiplet coupled to supergravity and an arbitrary number of chiral and vector multiplets.

The general \emph{on-shell} supergravity action for an arbitrary number of chiral and vector multiplets coupled to supergravity is given for example in equations (18.6)-(18.19) of the ``Supergravity'' book by Freedman and Van Proeyen \cite{freedman2012supergravity}. We would like to take one of the chiral multiplets to be nilpotent. However, the constraint equation \eqref{eq:s} given by $s=\frac{\bar \chi P_L \chi}{2 F}$ involves the auxiliary field $F$ that has already been integrated out in the on-shell supergravity action. Therefore one has to find a different starting point.

The standard approach is to start with a superconformal action and then do a gauge fixing and integrate out the auxiliary fields. This is described in detail for the standard supergravity action coupled to an arbitrary number of chiral and vector multiplets in chapter 16 to 18 of the book \cite{freedman2012supergravity}. In the presence of a nilpotent field one expects it to be difficult to integrate out the auxiliary field $F$ since it appears in the denominator of $s=\frac{\bar \chi P_L \chi}{2 F}$. Also since $s$ is a fermion bilinear the fermionic action is rather complicated. Nevertheless, it was possible, following this procedure, to derive the component action for the case of a nilpotent chiral superfield coupled to supergravity in \cite{Bergshoeff:2015tra, Hasegawa:2015bza}, and with one additional chiral multiplet in \cite{Hasegawa:2015bza}.\footnote{Since these component actions are rather long expressions, we refrain from spelling them out here and refer the interested reader to the original literature (see however equation \eqref{eq:LKW} and section \ref{sec:action} below).}

The action for one nilpotent chiral multiplet coupled to supergravity was termed \emph{pure dS supergravity} in \cite{Bergshoeff:2015tra} since it contains no scalar fields and allows for a positive cosmological constant. Previously no such action was known and there are no-go theorems \cite{Pilch:1984aw, Lukierski:1984it} that forbid such solutions in the case of linearly realized supersymmetry.

A shortcut in the derivation of the component action is to start directly from the supergravity action in which only the chiral superfields $Z^\alpha = z^\alpha +\sqrt{2}\theta \chi^\alpha + \theta^2 F^\alpha$ are taken off-shell. This action is given by \cite{Kallosh:2015tea} \footnote{Note that for standard chiral multiplets, one can simply integrate out $\bar{F}^{\bar \beta}$ from the action \eqref{eq:Loff} and find $(F^\alpha - F_G^\alpha)g_{\alpha \bar \beta}=0$. So in this case one trivially recovers the on-shell supergravity Lagrangian $\Ls{book}$.}
\be\label{eq:Loff}
e^{-1} \Ls{off-shell} = (F^\alpha - F_G^\alpha) g_{\alpha\bar \beta}(\bar{F}^{\bar{\beta}}-\bar{F}^{\bar \beta}_G) + e^{-1} \Ls{book}\,,
\ee
where 
\be\label{eq:FG}
F_G^\alpha = -{\rm e}^{\frac{K}{2}}g^{\alpha \bar\beta} \bar D_{\bar \beta} \bar W + {1\over2}\Gamma_{\beta \gamma}^\alpha \bar\chi^\beta \chi^\gamma +\frac14 \bar{f}_{AB\bar\beta}g^{ \alpha \bar \beta}\bar \lambda^A P_R \lambda^B\,.
\ee
The Lagrangian $\Ls{book}=\Ls{book}(e,\psi_\mu,z^\alpha,\chi^\alpha, \bar{z}^{\bar \alpha}, \chi^{\bar \alpha}, A^A_\mu, \lambda^A)$ is the general on-shell Lagrangian given in the book \cite{freedman2012supergravity} in equations (18.6)-(18.19). It depends on the vielbein, ${e_\mu}^a$, the gravitino, $\psi_\mu$, the physical fields of the chiral superfield, $Z^\alpha$, the vector fields, $A^A_\mu$ and the gauginos, $\lambda^A$. The action is determined through three functions of the scalar fields, $z^\alpha$, and their complex conjugates. These functions are the holomorphic superpotential, $W(z^\alpha)$, the holomorphic gauge-kinetic function, $f_{AB}(z^\alpha)$ and the real {\K} potential, $K(z^\alpha,\bar{z}^{\bar \alpha})$. We will also use $g_{\alpha \bar{\beta}} = \partial_\alpha \partial_{\bar \beta} K$ to denote the {\K} metric and $\Gamma^\alpha_{\beta \gamma}$ for the corresponding Christoffel symbols. Furthermore, we use subscripts as shorthand for partial derivatives, \emph{e.g.}: $K_\alpha =\partial_\alpha K$ and $f_{AB \alpha}=\partial_\alpha f_{AB}$. For our other conventions see the book \cite{freedman2012supergravity}. 

Having the off-shell action \eqref{eq:Loff} as starting point we can now demand that the first multiplet $Z^1\equiv S$ is nilpotent and split the index $\alpha=(1,i)$, with $i=2,3,\ldots,n$. This then means that $z^1 = {\bar{\chi}^1 P_L \chi^1 \over 2 F^1 } \equiv { (\chi^1)^2 \over 2 F^1 }$ so that now $\Ls{book}$ depends on $F^1$ via $z^1$ and we cannot simply integrate out $F^1$ anymore. It is however straight forward to integrate out the other $F^i$ which leads to 
\be
g_{i\b1}(\bar{F}^{\b1}-\bar{F}_G^{\b1} )+g_{i\jb}(\bar{F}^{\jb}-\bar{F}_G^{\jb})=0. 
\ee 
Since the sub-matrix $g_{i\jb}$ determines the kinetic terms for all the scalar fields (recall that $z^1$ is a fermion bilinear), it has to be invertible and we will denote its inverse by $(g_{i\jb})^{-1}$. This allows us to rewrite the above as
\be
(\bar{F}^{\jb}-\bar{F}_G^{\jb})=-(g_{i\jb})^{-1} g_{i\b1}(\bar{F}^{\b1}-\bar{F}_G^{\b1} )\,.
\ee
Plugging this back into the Lagrangian in equation \eqref{eq:Loff} we find 
\bea
e^{-1} { \cal L}_{\rm off-shell}&=& (F^1 - F^{1 } _G  ) \lp g_{1 \bar 1 } -g_{1\jb} (g_{\jb i})^{-1}  g_{i \bar 1} \rp   (\bar F^{\bar 1} - \bar F^{\bar 1 } _G) \cr
&&+ e^{-1} { \cal L}_{\rm book}\,.
\label{preoffshell}
\eea
Since $z^1$ and $\bar z^\b1$ square to zero we can also expand $\Ls{book}$ as follows
\be\label{eq:Lexp}
\Ls{book} = \bar z^\b1 A^1 z^1 +\bar B^1 z^1 +B^1 \bar z^\b1 + C^1\,.
\ee
The explicit expression for $A^1, B^1,\bar B^1, C^1$ are lengthy and can be read off from the explicit Lagrangian in the book \cite{freedman2012supergravity}. Note that $A^1$ contains derivatives that (after integration by parts) only act on $z^1$.

The discussion above follows \cite{Kallosh:2015sea, Kallosh:2015tea} and is the starting point for integrating out $F^1$ and deriving the explicit component action in the presence of the nilpotent chiral multiplet and an arbitrary number of chiral and vector multiplets. This was done in \cite{Kallosh:2015tea} under the assumption that the {\K} potential depends only on the product $z^1 \bar z^\b1$ and does not have linear terms in $z^1$ and $\bar z^\b1$. The resulting Lagrangian was surprisingly simple and is given by
\bea\label{eq:LKW}
e^{-1} \Ls{}&=& \ls e^{-1} \Ls{book} \rs_{z^1=\frac{(\chi^1)^2}{2 F^1_{G0}}, \bar z^\b1=\frac{(\chi^\b1)^2}{2 \bar F^\b1_{G0}}}\cr
&& - \frac{(\chi^1)^2 (\chi^\b1)^2}{4 g_{1\b1}\lp F^1_{G0} \bar{F}^\b1_{G0}\rp^2} \left|g_{1\b1}\frac{\square (\chi^1)^2}{2 F^1_{G0}} +B^1 \right|^2\,,
\eea
with $B^1$ defined in equation \eqref{eq:Lexp} and 
\be
F_{G0}^1 = \frac{1}{g_{1\b1}} \lp -{\rm e}^{\frac{K_0}{2}} \bar W_1 + \frac14 \bar{f}_{AB\b1}\bar \lambda^A P_R \lambda^B \rp\,.
\ee
Here $K_0(z^i,\bar{z}^\ib)$ is the $z^1 \bar z^{\b1}$ independent part of the {\K} potential $K(z^i,\bar{z}^\ib,z^1 \bar z^{\b1})$ and the indices on $W_1(z^i)$ and $\bar{f}_{AB\b1}(\bar z^\ib)$ denote partial derivatives as mentioned above.

So we see that in this case the action is given by the standard action where we take $z^1$ to be the fermion bilinear $(\chi^1)^2$ divided by $2 F^1_{G0}$ which is a function of the scalars $z^i,\bar z^\ib$ and the gauginos $\lambda^A$. Additionally, there is a new term that is quartic in the spinor $(\chi^1)^2 (\chi^\b1)^2$. This is the maximal power of the spinor and therefore we can set any terms in $B^1$ to zero that contain the undifferentiated spinor $\chi^1$ or $\chi^\b1$.

In the action above the goldstino is generically not just $\chi^1$ since for non-vanishing $F^i$ and/or non-vanishing D-terms the gravitino is a linear combination of $\chi^1$ and the $\chi^i$ and/or the $\lambda^A$.\footnote{$F^1$ cannot be zero since the nilpotent constraint is  $z^1 = { (\chi^1)^2 \over 2 F^1 }$, so that $\chi^1$ always appears in the goldstino.} A standard simplification in supergravity is to gauge fix the supersymmetry transformations to set the goldstino to zero (see section 5 of \cite{Kallosh:2015sea} for a detailed discussion of this point). This avoids mixing terms between the goldstino and the gravitino and simplifies the action $\Ls{book}$. However, due to large number of complicated terms involving $\chi^1$ it might be more useful to gauge fix the supersymmetry in equation \eqref{eq:VAtrafo} by setting $\chi^1=0$. In this case the above action \eqref{eq:LKW} reduces to 
\be\label{eq:Lchizero}
e^{-1} \Ls{}= \ls e^{-1} \Ls{book} \rs_{z^1=\bar z^\b1=\chi^1=\chi^\b1=0}\,.
\ee
This above action seems to be the standard action in the book. Note, however that, as mentioned above, we have already gauge fixed the supersymmetry transformations in such a way that $\Ls{book}$ contains generically mixing terms between the goldstino and the gravitino. Furthermore, despite the fact that $z^1=\chi^1=0$ the presence of the nilpotent field still leads to the positive definite contribution $K^{s\bar s} D_s W \overline{D_s W}$ in the scalar potential $V$ that appears in $\Ls{book}$ (see subsection \ref{sec:dS} above).

This concludes our review of the current literature. In the next section we will extend the results in \cite{Kallosh:2015tea} and derive the component action for the most general supergravity Lagrangian containing one nilpotent chiral multiplet and an arbitrary number of chiral and vector multiplets.

\section{The general dS supergravity action}\label{sec:action}
As in \cite{Kallosh:2015tea} we examine the supergravity Lagrangian for an arbitrary number of chiral and vector multiplets where one of the chiral superfields is nilpotent.  As above, we write the chiral superfields, $Z^\alpha = z^\alpha +\sqrt{2}\theta \chi^\alpha + \theta^2 F^\alpha$, and choose $Z^1\equiv S$ to be nilpotent and split the index $\alpha=(1,i)$, with $i=2,3,\ldots,n$.  The nilpotent constraint, $(Z^1)^2=0$, constrains the scalar $z^1$ to be a fermion bilinear given by $z^1 = {\bar{\chi}^1 P_L \chi^1 \over 2 F^1 } \equiv  { (\chi^1)^2 \over 2 F^1 }$. Then, the most general {\K} and superpotential and gauge kinetic functions are\footnote{In \cite{Kallosh:2015tea} the two functions $ K_1(z^i,\zb^\ib)$ and $K_\b1(z^i,\zb^\ib)$ where set to zero for simplicity.}
\bea
W(z^\alpha) &=& W_0(z^i) + W_1(z^i) z^1\,,\cr
K(z^\alpha,\zb^{\bar \alpha}) &=& K_0(z^i,\zb^\ib) + K_1(z^i,\zb^\ib)\ z^1 + K_\b1(z^i,\zb^\ib)\ \zb^\b1 \cr
&&+ g_{1\b1}(z^i,\zb^\ib) \ z^1 \zb^\b1\,,\cr
f_{AB}(z^\alpha) &=&f_{AB0}(z^i) + f_{AB1}(z^i)\ z^1\,,
\eea
where $\partial_{\alpha \bar \beta} K = g_{\alpha \bar \beta}$. 

As explained above, since $z^1={(\chi^1)^2 \over 2 F^1 }$ depends on $F^1$, we need to start with the off-shell supergravity Lagrangian, and then proceed to integrate out $F^1$ (taking into account its appearance in $z^1$). Following \cite{Kallosh:2015tea}, we take the off-shell Lagrangian in which all the $F^\alpha$ auxiliary fields are not yet integrated out. Then it is trivial to integrate out the $F^i$, as shown in the previous subsection. This leads to the Lagrangian in equation \eqref{preoffshell} that is off-shell with respect to $F^1$ only:
\be \label{offshell}
\begin{split}
e^{-1} { \cal L}_{\rm off-shell}=& (F^1 - F^{1 } _G  ) \lp g_{1 \bar 1 } -g_{1\jb} (g_{\jb i})^{-1}  g_{i \bar 1} \rp   (\bar F^{\bar 1} - \bar F^{\bar 1 } _G) \\
&+ e^{-1} { \cal L}_{\rm book}\,.
\end{split}
\ee
It will be convenient for us to define
\be
\begin{split}
g_{1 \bar 1 } -g_{1\jb} (g_{\jb i})^{-1}  g_{i \bar 1} &= M_0 + M_1 z^1 + M_{\bar 1} \bar{z}^{\bar 1} + M_{1\bar1} z^1 \bar{z}^{\bar1}\\
&\equiv \mathcal{M}\,,
\end{split}
\ee
where the quantities $M_{0,1,\bar{1},1\bar1}$ are given explicitly in the appendix in equation (\ref{Mdef}).
Similarly we can expand $F^1_G$ and $\bar F^\b1_G$  as
\bea \label{fg1def}
F_G^1&=& F_{G0}^1 + F_{G1}^1z^1+F_{G\bar 1}^1\bar z^{\bar 1} + F_{G1\bar1}^1z^1\bar z^{\bar1}\,,\cr
\bar{F}_G^\b1&=& \bar F_{G0}^\b1 + \bar F_{G1}^\b1z^1+\bar F_{G\bar 1}^\b1\bar z^{\bar 1} + \bar F_{G1\bar1}^1z^1\bar z^{\bar1}\,.
\eea
We can absorb all the terms of $F^1_G$ that depends on $z^1$ into the following redefinition of $F^1$
\begin{equation}
F'^1 \equiv F^1- F_{G1}^1z^1- F_{G1\bar1}^1z^1\bar z^{\bar1}.
\label{Fprimedef}
\end{equation}
This definition preserves the form of the nilpotency condition on $z^1$, giving $z^1 = {(\chi^1)^2 \over 2 F^1 } =  { (\chi^1)^2 \over 2 F'^1 }$, since $(\chi^1)^2z^1 \propto (\chi^1)^2(\chi^1)^2=0$. Likewise we define $\bar{F}'^{\bar{1}}$ as the complex conjugate of \eqref{Fprimedef}. We can rewrite the first term of (\ref{offshell}) as
\be \label{auxterm}
\begin{split}
&(F^{'1} -F^1_{G0}- F^{1 } _{G\b1} \zb^\b1  )\mathcal{M}   (\bar F^{'\bar 1} -\bar F^\b1_{G0} - \bar F^{\bar 1 } _{G1} z^1)\,.
\end{split}
\ee
To integrate out $F'^{1}$ it will be useful to write the Lagrangian in the following form
\begin{align}
e^{-1} &{ \cal L}_{\rm off-shell}= (F^{'1} -F^1_{G0}) M_0  (\bar F^{'\bar 1} -\bar F^\b1_{G0}) \nonumber \\
& + \frac{1}{2} (\chi^1)^2 M_1 (\bar F^{'\bar 1} -\bar F^\b1_{G0}) - z^1 F^1_{G0} M_1 (\bar F^{'\bar 1} -\bar F^\b1_{G0}) \nonumber \\
& + \frac{1}{2} (\chi^{\bar{1}})^2 (F^{'1} -F^1_{G0}) M_{\bar{1}} - \zbar (F^{'1} -F^1_{G0}) M_{\bar{1}} \bar{F}^{\bar{1}}_{G0} \nonumber \\
&+ \zbar A z^1 + B \zbar + \bar{B} z^1 + C\,.
\label{offshellexplicit}
\end{align}
In this form the explicit dependence of the Lagrangian on $F'^{1}$ and $z^1$ is apparent and we used that the product $F'^{1}z^1=\frac12 (\chi^1)^2$ is independent of $F'^{1}$. The coefficients in the last line are
\begin{align}
A = &A^1+ F_{G0}^1 M_{1\bar{1}} \bar{F}^{\bar{1}}_{G0}  + F_{G0}^1 M_{\bar{1}} \bar{F}^{\bar{1}}_{G1} + F_{G\bar{1}}^1 M_{1} \bar{F}^{\bar{1}}_{G0} \nonumber \\
& \quad  + F^1_{G\bar{1}} M_0 \bar{F}^{\bar{1}}_{G1}, \\
B = &B^1  - \frac{1}{2} (\chi^1)^2 \left[M_{1\bar{1}} \bar{F}^{\bar{1}}_{G0} + M_{\bar{1}} \bar{F}^{\bar{1}}_{G1}  \right] + F^1_{G\bar{1}}M_0 \bar{F}^{\bar{1}}_{G0}, \\
\bar{B} = &\bar{B}^1  - \frac{1}{2} (\chi^{\bar{1}})^2 \left[M_{1\bar{1}} {F}^{{1}}_{G0} + {F}^{{1}}_{G\bar{1}} M_{{1}}   \right] + F^1_{G0} M_0 \bar{F}^{\bar{1}}_{G1}, \\
C = &C^1+ \frac{1}{4} (\chi^{\bar{1}})^2(\chi^1)^2 M_{1\bar{1}}  - \frac{(\chi^1)^2}{2} M_0 \bar{F}^{\bar{1}}_{G1} \nonumber \\
& \quad -  \frac{(\chi^{\bar{1}})^2}{2} M_0F^{1}_{G\bar{1}}\,,
\end{align}
where $A^1, B^1, \bar B^1, C^1$ are defined via the expansion in equation \eqref{eq:Lexp}.

We can now follow the method developed in \cite{Bergshoeff:2015tra, Hasegawa:2015bza} to integrate out $F'^1$. We need to solve the field equation for $F'^1$, which is given by
\begin{align}
0 =& \frac{\partial \mathcal{L}}{\partial \bar{F}'^{\bar{1}}} - \frac{\zbar}{\bar{F}'^{\bar{1}}} \frac{\partial \mathcal{L}}{\partial \bar{z}^{\bar{1}}} \nonumber \\
 =& (F^{'1} -F^1_{G0}) M_0 + z^1 (F^{'1} -F^1_{G0}) M_1 \nonumber \\
&\quad - \frac{\zbar}{\bar{F}'^{\bar{1}}} \left[-(F^{'1} -F^1_{G0}) M_{\bar{1}} \bar{F}^{\bar{1}}_{G0} + A z^1 + B \right].
\label{Feom}
\end{align}
This equation can be solved as in \cite{Hasegawa:2015bza} by inserting the expansion
\be
F^{'1}=F^{'1}_0 + F^{'1}_1 (\chi^1)^2 +F^{'1}_\b1  (\chi^\b1)^2 + F^{'1}_{1\b1} (\chi^1)^2(\chi^\b1)^2,
\ee
and its complex conjugate, into \eqref{Feom}. The nilpotency of $(\chi^1)^2$ and $(\chi^\b1)^2$ allows us to do an exact Taylor expansion where $F^{'1}$ (or its complex conjugate) appears in the denominator.  We then see that there are four equations for the four unknown $F^{'1}_0$, $F^{'1}_1$, $F^{'1}_\b1$ and $F^{'1}_{1\b1}$. When doing this one has to recall that $A$ contains two derivatives so that $A z^1$ contains a term proportional to $\square (\chi^1)^2$. This means in particular that $(\chi^1)^2 A z^1 \neq 0$. Solving \eqref{Feom} order by order we find the result
\begin{align}
F'^1 = &F^1_{G0} + \frac{(\chi^{\bar{1}})^2}{2 (\bar{F}^{\bar{1}}_{G0})^2 M_0} \lp A \frac{(
\chi^1)^2}{2F_{G0}^1} + B \rp \nonumber \\
& \quad \times \left[1- \frac{(
\chi^1)^2}{F_{G0}^1 M_0} \left(\frac{M_1}{2} + \frac{\lp \bar{A} \frac{(
\chi^\b1)^2}{2\bar F_{G0}^1} + \bar{B}\rp }{F^1_{G0}\bar{F}^{\bar{1}}_{G0}} \right) \right] \,.
\label{Fsolution}
\end{align}
Note that $(F'^1 - F^1_{G0})$ is proportional to $(\chi^{\bar{1}})^2$ (and likewise $(\bar F'^{\b1} - \bar F^{\b1}_{G0}) \propto (\chi^1)^2$). This means that upon plugging this into the Lagrangian \eqref{offshellexplicit} we find that the on-shell Lagrangian will simplify significantly. In particular, for the term in equation \eqref{auxterm} one can replace $\mathcal{M}$ with $M_0$.

When we plug the solution \eqref{Fsolution} and its complex conjugate back into the Lagrangian \eqref{offshellexplicit} we find 
\begin{align}
e^{-1} &\mathcal{L}_{\text{on-shell}} = \frac{(\chi^{\bar{1}})^2}{2 \bar{F}^{\bar{1}}_{G0}} A \frac{(\chi^1)^2}{2 F^1_{G0}} + B \frac{(\chi^{\bar{1}})^2}{2 \bar{F}^{\bar{1}}_{G0}} + \bar{B} \frac{(\chi^1)^2}{2 F^1_{G0}} \nonumber + C\\
&  - \frac{(\chi^{\bar{1}})^2 (\chi^1)^2}{4 (\bar{F}^{\bar{1}}_{G0})^2 (F^1_{G0})^2 M_0} 
\Bigg| A \frac{(\chi^1)^2}{2 F^1_{G0}} + B \Bigg|^2,
\label{onshell1}
\end{align} 
If we plug in the explicit expressions for $A,B,\bar B$ and $C$, we find that all terms in the first line of \eqref{onshell1} that depend on $M_1$, $M_{\bar{1}}$ and $M_{1\bar{1}}$ cancel with each other. Moreover, one can show that these terms also vanish in the last line of \eqref{onshell1} due to the fact that $(\chi^1)^4 =0$. The only terms in $A$ that survive are the ones that contain derivatives that act on $(\chi^1)^2$. We also find that all terms in the last three terms of the first line of \eqref{onshell1} that depend on $M_0$ cancel. So finally we are left with:
\begin{align} \label{L2}
&e^{-1} \mathcal{L}_{\rm on-shell} =  \ls  e^{-1} \mathcal{L}_{\rm book}\rs_{z^1=\frac{(\chi^1)^2}{2 F^1_{G0}}} + \frac{(\chi^{\bar{1}})^2(\chi^1)^2}{4 F^1_{G0} \bar{F}^{\bar{1}}_{G0}}  \\
&\times \ls F^1_{G\bar{1}} M_{0} \bar{F}^{\bar{1}}_{G1} - \frac{1}{M_{0} F^1_{G0} \bar{F}^{\bar{1}}_{G0}} \left|  \frac{A^1 (\chi^1)^2}{2 F_{G0}^1} +  B^1 + F^1_{G\bar{1}} M_{0} \bar{F}^{\bar{1}}_{G0} \right|^2 \rs.\nn
\end{align}
Note that the last term in the square will exactly cancel the first term in the square brackets, so the only additional terms will come from the cross terms of $A^1$ and $B^1$ with $\bar{F}^{\bar{1}}_{G1} M_0 F^1_{G0}$ and its conjugate. 
The explicit forms for $F_{G0}^1$ and $F_{G\bar1}^1$ are given in Appendix \ref{appendixC}.

\section{Conclusion} \label{sec:conclusion}

In this paper we reviewed the appearance and utility of a nilpotent chiral multiplet in $4d$ $\mathcal{N}=1$ supergravity. One of its most important features is that the nilpotent chiral multiplet coupled to supergravity allows one to evade no-go theorems and construct dS solutions \cite{Antoniadis:2014oya} (even in the absence of any scalar degrees of freedom), hence the resulting theory is called \emph{pure dS supergravity} \cite{Bergshoeff:2015tra}. This feature makes the nilpotent chiral multiplet a very useful ingredient in cosmological model building in supergravity. In this context the nilpotent chiral multiplet does not only lead to models that can accommodate a dS vacuum and hence the accelerated expansion of our current universe, but it also simplifies inflationary model building in many cases \cite{Ferrara:2014kva}.

We also reviewed how the nilpotent chiral multiplet is connected to dS vacua in string theory that rely on an anti-D3-brane uplift. There one can explicitly show that the action for the anti-D3-brane in a flux background reduces to the Volkov-Akulov action for the goldstino. Since branes and anti-branes break supersymmetry spontaneously we expect that there are many more string theory construction that one can connect to the nilpotent chiral multiplet.

Since the scalar component in the nilpotent chiral multiplet is a fermion bilinear one rightfully expects that its fermionic action is rather complicated. Nevertheless, it was possible in \cite{Bergshoeff:2015tra, Hasegawa:2015bza, Kallosh:2015sea, Kallosh:2015tea} to derive the explicit component action for a nilpotent chiral multiplet coupled to supergravity in a variety of different examples. In particular, in \cite{Kallosh:2015tea} the action for a nilpotent chiral multiplet coupled to supergravity and an arbitrary number of chiral and vector multiplets was derived under the single simplifying assumption that the scalar component of the nilpotent chiral multiplet, $s$, only appears in the combination $s\bar{s}$ in the {\K} potential. Here we dropped this assumption and we were able to derive the most general action for a nilpotent chiral multiplet coupled to supergravity and an arbitrary number of chiral and vector multiplets. Our result is given in equation \eqref{L2} in section \ref{sec:action}.
 
Our resulting action can be simplified by gauge fixing the supersymmetry transformations. In particular in the gauge $\chi=0$ the action becomes rather simple (cf. eqn. \eqref{L2}). Note however that although in this gauge $s\equiv z^1 = \chi^1=0$ there is still an important contribution from the nilpotent chiral multiplet in the action since $F^1 = F^1_{G0} \neq 0$. In this particular gauge the gravitino is generally not zero, so that there are many higher order fermionic couplings (see section 5 in \cite{Kallosh:2015sea} for a detailed discussion of this point). For that reason it is very useful to have the entire component action before gauge fixing so that one can choose the simplest gauge for any given problem.

The entire component action in \eqref{L2} including all the fermionic terms is of great importance. In the cosmological context, which is the focus of this paper, one has to reheat the universe at the end of inflation and this requires the detailed knowledge of all the couplings in the model. In particular, if we want to couple a given cosmological model of inflation that uses the nilpotent chiral multiplet to the standard model of particle physics, then we of course need to know the action including chiral and vector multiplets. It should also be interesting to study the action \eqref{L2} purely from a particle physics point of view.

\section*{Acknowledgments}

We are grateful to S. Ferrara, D. Freedman,  R. Kallosh and A. Van Proeyen for stimulating  discussions. We thank S. Ferrara, R. Kallosh and A. Van Proeyen for valuable comments on a draft of this paper. The work of MS is supported by the European Union's Horizon 2020 research and innovation programme under the Marie Sklodowska-Curie grant agreement No 656491. The work of EW is supported by the National Science Foundation of Belgium (FWO) grant G.001.12 Odysseus. We thank the organizers of the conference ``The String Theory Universe, 21st European string workshop and 3rd COST MP1210 meeting'' for a wonderful workshop during which this work was initiated.

\appendix

\section{Inverse K\"ahler  metric and Christoffel symbols} \label{appendixB}
Here we collect the expressions for the inverted K\"ahler metric and the Christoffel symbols.  The K\"ahler metric is:
\be
\begin{split}
g_{1\bar{1}} &= g_{1\bar{1}} \\
g_{1\ib} &= K_{1,\ib} + \zbar g_{1\bar{1},\ib} \\
g_{i\bar{1}} &= K_{\bar{1},i} + z^1 g_{1\bar{1},i} \\
g_{i\jb} &= K_{0,i\jb} + z^1 K_{1,i\jb} + \zbar K_{\bar{1},i\jb} + z^1 \zbar g_{1\bar{1},i\jb}.
\end{split}
\ee
Inverting the matrix $g_{i\jb}$, one finds
\begin{align}
&(g_{\jb i})^{-1} = \kinv{j}{i} - z^1\kinv{j}{l} K_{1,l\bar{m}} \kinv{m}{i} \cr
&\quad - \zbar \kinv{j}{l} K_{\bar{1},l\bar{m}}  \kinv{m}{i} + z^1 \zbar \kinv{j}{l}\left[- \gone{,l\bar{m}}  \right. \cr
&\quad \left. + K_{1,l\bar{t}} \kinv{t}{s} K_{\bar{1},s\bar{m}} + K_{\bar{1},l\bar{t}} \kinv{t}{s} K_{1,s\bar{m}}\right]\kinv{m}{i} \cr
&\equiv \kinv{j}{i} + z^1 D_{\jb i} + \zbar E_{\jb i} + z^1\zbar F_{\jb i} .
\end{align}
We will often need the combination,
\be \label{Mdef}
\begin{split}
&g_{1\bar1} - g_{1\jb}(g_{\jb i})^{-1}  g_{i\bar{1}}= g_{1\bar1} - K_{1,\jb} \kinv{j}{i} K_{\bar{1},i} \\
&\quad - z^1  K_{1,\jb} \left[ D_{\jb i} K_{\bar{1},i} +  \kinv{j}{i} \gone{,i} \right] \\
&\quad- \zbar  \left[ K_{1,\jb} E_{\jb i}  +   \gone{,\jb} \kinv{j}{i} \right] K_{\bar{1},i}  \\
&\quad - z^1 \zbar \left [ K_{1,\jb} F_{\jb i} K_{\bar{1},i} + K_{1,\jb} E_{\jb i} \gone{,i} + \gone{,\jb} D_{\jb i} K_{\bar{1},i} \right. \\
&\quad \left. + \gone{,\jb} \kinv{j}{i} \gone{,i}  \right] \\
\equiv&  M_0 + M_1 z^1 + M_{\bar 1} \bar{z}^{\bar 1} + M_{1\bar1} z^1 \bar{z}^{\bar1}.
\end{split}
\ee
Now we compute the necessary components of the inverse metric:
\be
\begin{split}
g^{1\bar{1}} &= \lp g_{1\bar1} - g_{1\jb}(g_{\jb i})^{-1}  g_{i\bar{1}} \rp ^{-1} \\
&= {1 \over M_0} - z^1 {M_1 \over M_0^2} - \zbar {M_{\bar1} \over M_0^2} - z^1\zbar \lp {M_{1\bar1} \over M_0^2} -2 {M_1M_{\bar1} \over M_0^3}\rp \\
&\equiv (g^{1\bar{1}})_{(0)} + z^1 (g^{1\bar{1}})_{(1)} + \zbar (g^{1\bar{1}})_{(\bar{1})} + z^1 \zbar (g^{1\bar{1}})_{(1\bar{1})}.
\end{split}
\ee
\begin{align}
g^{1{\jb}} =&-(g_{\jb i})^{-1}g_{i\bar1}g^{1\bar1} \cr
=&  - \frac{\kinv{j}{i} K_{\bar{1},i}}{M_0}  
+ z^1 \left(-\frac{D_{\jb i} K_{\bar{1},i}}{M_0} + \frac{\kinv{j}{i} K_{\bar{1},i} M_1}{M_0^2} \right. \cr
& \left. - \frac{\kinv{j}{i} \gone{,i}}{M_0}\right) 
+ \zbar \left(-\frac{E_{\jb i} K_{\bar{1},i}}{M_0} + \frac{\kinv{j}{i} K_{\bar{1},i}  M_{\bar1}  }{M_0^2} \right)\cr
&+ z^1 \zbar  \Bigg[ - \frac{F_{\jb i} K_{\bar{1},i}}{M_0} - \frac{E_{\jb i} \gone{,i}}{M_0} + \frac{E_{\jb i} K_{\bar{1},i} M_1}{M_0^2} \cr
& + \frac{D_{\jb i} K_{\bar{1},i} M_{\bar1}  }{M_0^2} - \kinv{j}{i} K_{\bar{1},i} \lp2 {  M_1 M_{\bar1} \over M_0^3} - {M_{1\bar1} \over M_0^2} \rp  \cr
&+ \frac{\kinv{j}{i} \gone{,i} M_{\bar1} }{M_0^2}  \Bigg] \\
\equiv& (g^{1\jb})_{(0)} + z^1 (g^{1\jb})_{(1)} + \zbar (g^{1\jb})_{(\bar{1})} + z^1 \zbar (g^{1\jb})_{(1\bar{1})}.\nn
\end{align}
With these components we now calculate the Christoffel symbols:
\allowdisplaybreaks[1]
\begin{align}
\Gamma^1_{11} =& 0, \nn\\
\Gamma^1_{1i} =& (g^{1\bar{1}})_{(0)} \gone{,i} + (g^{1\jb})_{(0)} K_{1,i\jb} + z^1 \left[(g^{1\bar{1}})_{(1)} \gone{,i} \right. \nonumber\\
+&\left.  (g^{1\jb})_{(1)} K_{1,i\jb} \right]  + \zbar  \left [ (g^{1\bar{1}})_{(\bar{1})} \gone{,i}+ (g^{1\jb})_{(\bar{1})} K_{1,i\jb} \right. \nonumber \\
+& \left. (g^{1\jb})_{(0)} \gone{,i\jb} \right] + z^1 \zbar \left[(g^{1\bar{1}})_{(1\bar{1})} \gone{,i}   \right. \nonumber \\
+&\left.  (g^{1\jb})_{(1\bar{1})} K_{1,i\jb}+ (g^{1\jb})_{(1)} \gone{,i\jb} \right]\\
\equiv& \lp \Gamma^1_{1i}\rp_{(0)} + z^1 \lp\Gamma^1_{1i}\rp_{(1)} + \zbar \lp \Gamma^1_{1i}\rp_{(\bar{1})} + z^1 \zbar \lp\Gamma^1_{1i}\rp_{(1\bar{1})}\,,\nn\\
\Gamma^1_{ij} =& (g^{1\bar{1}})_{(0)} K_{\bar{1},ij} +  (g^{1\bar{k}})_{(0)} K_{0,ij\bar{k}} + z^1 \left[(g^{1\bar{1}})_{(1)} K_{\bar{1},ij} \right. \nonumber \\
+&\left. (g^{1\bar{1}})_{(0)} \gone{,ij}  +  (g^{1\bar{k}})_{(1)} K_{0,ij\bar{k}}  +   (g^{1\bar{k}})_{(0)} K_{1,ij\bar{k}} \right] \nonumber \\
+& \zbar \left[(g^{1\bar{1}})_{(\bar{1})} K_{\bar{1},ij} +   (g^{1\bar{k}})_{(\bar{1})} K_{0,ij\bar{k}} +  (g^{1\bar{k}})_{(0)} K_{\bar{1},ij\bar{k}} \right]  \nonumber \\
+& z^1 \zbar \left[(g^{1\bar{1}})_{(1\bar{1})} K_{\bar{1},ij} + (g^{1\bar{1}})_{(\bar{1})} \gone{,ij}  +  (g^{1\bar{k}})_{(1\bar{1})} K_{0,ij\bar{k}} \right. \nonumber \\
+&  \left. (g^{1\bar{k}})_{(\bar{1})} K_{1,ij\bar{k}} +  (g^{1\bar{k}})_{(1)} K_{\bar{1},ij\bar{k}} +   (g^{1\bar{k}})_{(0)} \gone{,ij\bar{k}} \right]\nn\\
\equiv&\lp \Gamma^1_{ij}\rp_{(0)} + z^1 \lp \Gamma^1_{ij}\rp_{(1)} + \zbar \lp\Gamma^1_{ij}\rp_{(\bar{1})} + z^1 \zbar \lp \Gamma^1_{ij}\rp_{(1\bar{1})}\,.\nn
\end{align}

\section{The expansion of $F_G^1$} \label{appendixC}
Using the expressions in Appendix \ref{appendixB} for the inverse metric and Christoffel symbols we can expand the definition of $F_G^1$ (see equation \eqref{eq:FG}):
\be 
F_G^1 = -{\rm e}^{K/2}g^{1\bar\beta} \bar D_{\bar \beta} \bar W + {1\over2}\Gamma_{\beta \gamma}^1 \bar\chi^\beta \chi^\gamma +\frac14 \bar{f}_{AB\bar\beta}g^{\bar \beta 1}\bar \lambda^A P_R \lambda^B,
\ee
to find $F_{G0}^1$ and $F_{G\bar1}^1$ as defined in equation \eqref{fg1def}.  We find:
\be
\begin{split}
F^1_{G0} = &\frac{1}{M_0} \Bigg[- e^{K_0/2} \Big(\bar{W}_{\bar{1}} + K_{\b1} \bar{W}_0 - \kinv{i}{j}K_{\bar{1},j}\left[\bar{W}_{0,\ib} \right. \\
&\left.+ K_{0,\ib} \bar{W}_0 \right] + \Big( \gone{,i}  - K_{1,i\bar{k}} \kinv{k}{j} K_{\bar{1},j}  \Big) \frac{\bar{\chi}^1\chi^i}{2} \\
&+ \Big( K_{\bar{1},ij}  - K_{0,ij\bar{k}} \kinv{k}{m} K_{\bar{1},m}  \Big) \frac{\bar{\chi}^i\chi^j}{2}  \\
& + \frac{1}{4} \left(\bar{f}_{AB\bar{1}} - \bar{f}_{AB\ib} \kinv{i}{j} K_{\b1,j} \right) \bar\lambda^A P_R \lambda^B \Bigg]\,,
\end{split}
\ee
and
\begin{align}
F^{1}_{G\bar{1}} = & -e^{K_0/2} \Bigg[\frac{1}{2}K_{\bar{1}} \left( (g^{1\bar{1}})_{(0)} (\bar{W}_{\bar{1}} + K_{\b1} \bar{W}_0) + (g^{1\ib})_{(0)} (\bar{W}_{0,\ib} \right. \nonumber \\
+&\left. K_{0,\ib} \bar{W}_0) \right) + (g^{1\bar{1}})_{(\bar{1})} (\bar{W}_{\bar{1}} + K_{\b1} \bar{W}_0) + (g^{1\bar{1}})_{(0)}K_{\b1} \bar{W}_{\bar{1}} \nonumber \\
+& (g^{1\ib})_{(\bar{1})} (\bar{W}_{0,\ib} + K_{0,\ib} \bar{W}_0) \cr
+& (g^{1\ib})_{(0)} (\bar{W}_{\b1 \ib} + K_{\bar{1},\ib}\bar{W}_{0} + K_{0,\ib} \bar{W}_{\bar{1}}) \Bigg] \nonumber \\
+& \left(\Gamma^1_{1i} \right)_{(\bar{1})} \frac{\bar{\chi}^1\chi^i}{2} + \left(\Gamma^1_{ij} \right)_{(\bar{1})} \frac{\bar{\chi}^i\chi^j}{2} + \frac{1}{4}\left((g^{1\bar{1}})_{(\bar{1})}\bar{f}_{AB\bar{1}} \right. \nonumber \\
+&\left. (g^{1\ib})_{(0)}\bar{f}_{AB\b1\ib} +(g^{1\ib})_{(\bar{1})}\bar{f}_{AB\ib} \right)\lambda^A P_R \lambda^B\,.
\end{align}

\bibliographystyle{JHEP} 
\bibliography{refs}

\end{document}